# Validation of the rapid detection approach for enhancing the electronic nose systems performance, using different deep learning models and support vector machines

Juan C. Rodriguez Gamboa[a,*], Adenilton J. da Silva[b], Ismael C. S. Araujo[c], Eva Susana Albarracin E.[a], *Cristhian M. Duran A.*[d]

[a] *Departamento de Estatística e Informática, Universidade Federal Rural de Pernambuco - UFRPE, Recife, PE, Brazil*

[b] *Centro de Informática, Universidade Federal de Pernambuco - UFPE, Recife, PE, Brazil*

[c] *Departamento de Computação, Universidade Federal Rural de Pernambuco, Recife, PE, Brazil*

[d] *Facultad de ingeniería y arquitectura, Universidad de Pamplona, Pamplona, NdeS, Colombia*

[*] *Corresponding author. Departamento de Estatística e Informática, Universidade Federal Rural de Pernambuco - UFRPE, Rua Dom Manoel de Medeiros s/n, Dois Irmãos, Recife, PE, Brazil. E-mail: juan.gamboa@ufrpe.br*

**Abstract**

Real-time gas classification is an essential issue and challenge in applications such as food and beverage quality control, accident prevention in industrial environments, for instance. In recent years, the Deep Learning (DL) models have shown great potential to classify and forecast data in diverse problems, even in the electronic nose (E-Nose) field. In this work, we used a Support Vector Machines (SVM) algorithm and three different DL models to validate the rapid detection approach (based on processing an early portion of raw signals and a rising window protocol) over different measurement conditions. We performed a set of trials with five different E-Nose databases that include fifteen datasets. Based on the results, we concluded that the proposed approach has a high potential, and it can be suitable to be used for E-nose technologies, reducing the necessary time for making forecasts and accelerating the response time. Because in most cases, it achieved reliable estimates using only the first 30% or fewer of measurement data (counted after the gas injection starts.) The findings suggest that the rapid detection approach generates reliable forecasting models using different classification methods. Still, SVM seems to obtain the best accuracy, right window size, and better training time.

## 1. Introduction

The conventional approach for data processing in the Electronic Nose implies using the entire response curves (including rising, steady-state, recovery phases, and other) of the gas sensors array. Besides, this approach includes steps such as signal preprocessing and feature generation/extraction before performing the classification tasks, which requires the selection of a suitable method for each stage, increasing the necessary time to find the appropriate classification and forecasting models [1,2]. Nowadays, some researches have focused their efforts on reducing the steps and the essential know-how for model generation, such as the works presented by [3] and [4]. In the first case, the authors proposed a bio-inspired data processing method based on a neural network to mimic the mammalian olfactory system with excellent results but using the entire measurement curves. In the second case, the authors proposed a rapid detection system for meat spoilage using an unsupervised technique (i.e., stacked restricted Boltzmann machines and auto-encoders) that considers only the transient response. Although the obtained models offer advantages because the features are learned from data instead of being hand-designed, it may produce low suitable and inaccurate models due to the unsupervised method. Furthermore, some authors have explored an approach based on raw data treatment [5,6]. Although this approach reduces the steps and the development time, they only tested with the entire response curves, requiring to complete all measurement processes. So, it can take critical time to get results.

Concerning the above mentioned, we proposed a novel approach on [7], based on processing an early portion of signals (while the measurement process is still running.) We also tested the proposed method in a wine quality application, with excellent results against the traditional methodology. A deep MLP classifier was trained with the raw data acquired from an E-Nose composed of an array of six MOX gas sensors. We achieved results around 63 times faster (Eq. 1) compared with a traditional method (using the entire response curves, applying preprocessing techniques to extract the features and later processing them using an SVM algorithm.)

$$relation\ of\ measurement\ time = \left(\frac{measurement\ time\ from\ the\ starting\ gas\ injection\ to\ the\ finish}{necessary\ time\ for\ making\ a\ forecast\ or\ window\ size}\right) \quad \text{Eq. (1)}$$

Support Vector Machines (SVM) is one of the most applied methods for classification in E-Nose. Other used methods are K-Nearest Neighbors (KNN), Naive Bayes (NB), Linear Discriminant Analysis (LDA), and Adaptive Resonance Theory Map (ARTMAP) [8]. However, the more novel approaches are based on neural networks, implementing variations about the architectures, learning techniques, for instance. In recent years the deep learning algorithms have arrived at the electronic nose field, such as the case of [1,2,5,6], where the authors have explored the Convolutional Neural Networks (CNN) in different settings.

The present study focuses on validating if the rapid detection approach is suitable to be applied in diverse E-Nose settings (five different databases) [7]. Additionally, to test DL techniques such as Convolutional Neural Networks (CNN) against a more classical method like SVM for classification tasks in E-Nose using the proposed approach.

2. **Materials and methods**

In this work, we used five E-Nose databases that include fifteen datasets to test our approach. The tested databases correspond to different systems used in several settings and with distinct experimental setups, guaranteeing varied conditions. We intend to make this work as a reference for further researches, encouraging the research community to perform more studies or analysis by using E-Noses with numerous databases as well as making public the collected databases.

*2.1 Databases*

*2.1.1 Database 1: Electronic nose dataset for the detection of wine spoilage thresholds*

This public database consists of time series collected through an electronic nose for a wine quality control application focused on spoilage thresholds. This database has two datasets, one of them composed of only wines (three-class classification problem), and the other comprises wines and ethanol (four classes). The database contains 235 recorded measurements of wines divided into three groups, labeled as high quality (HQ), average quality (AQ), and low quality (LQ), in addition to 65 ethanol measurements. The time series acquired at 18.5 Hz of sampling frequency during 180 seconds correspond to 3330 data points per sensor. Each file in the dataset has eight columns: relative humidity (%), temperature (°C), and the resistance readings in kΩ of the six MOX gas sensors: MQ-3, MQ-4, MQ-6, MQ-3, MQ-4, MQ-6. More details are available in [7,9].

*2.1.2 Database 2: Electronic Nose for Quality Control of Colombian Coffee through the Detection of Defects in "Cup Tests"*

This dataset consists of time-series recorded by an electronic nose used for the coffee quality control to detect defects in the grain [10,11]. The dataset contains 58 measurements of coffee divided into three groups and labeled as high quality (HQ), average quality (AQ), and low quality (LQ), inducing a three classes classification problem. The time series acquired at 1 Hz of sampling frequency during 300 seconds correspond to 2400 data points for each measurement. Where, each file in the dataset has eight columns with the resistance readings in kΩ of the gas sensors: SP-12A, SP-31, TGS-813, TGS-842, SP-AQ3, TGS-823, ST-31, TGS-800.

*2.1.3 Database 3: Gas sensor arrays in open sampling settings Data Set*

The authors compiled an extensive database through a chemical detection platform for detecting potentially hazardous substances at different concentrations, composed of nine portable sensor array modules (72 metal-oxide chemical sensors in a wind tunnel facility.) Each module had eight MOX gas sensors (manufactured by Figaro Inc.) and positioned at six different line locations normal to the wind direction. Thus, creating thereby a total number of 54 measurement locations, uniformly distributed for a total of 18000 different measurements. We split this database into six datasets (each dataset corresponds to one line location: L1, L2, …, L6.) Compounds, such as acetone, acetaldehyde, ammonia, butanol (butyl-alcohol), ethylene, methane, methanol, carbon monoxide, benzene, and toluene (ten classes) were measured to generate the database [12].

*2.1.4 Database 4: Gas sensor array exposed to turbulent gas mixtures Data Set*

This dataset was obtained using the same wind tunnel mentioned in section 2.1.3, but the wind tunnel was adapted from the previous setup to include two independent gas sources. Besides, only one module (eight MOX gas sensors array) was used in a fixed location in the wind tunnel. The sensors array was exposed to binary mixtures of ethylene with either methane or carbon monoxide. Volatile Organic Compounds (VOCs) were released at four different rates to induce different concentration levels in the module vicinity. Each configuration was repeated six times, for a total of 180 measurements. See [13,14] for additional details. In this work, we split the dataset to generate a four-class classification problem, including the followings categories (high ethylene concentration, medium ethylene concentration, low ethylene concentration, and without ethylene.) Hence, this is a challenging problem because the measurements were performed using two interfering gases (Methane, carbon monoxide) at different concentrations, and all groups of measurements include binary mixtures of ethylene with combinations of the mentioned interfering VOC.

*2.1.5 Database 5: Twin gas sensor arrays Data Set*

This database comprises the recordings of five twin eight gas sensor detection units. This database has five datasets (B1, B2, …, B5) where each dataset corresponds to the measurements of one twin system (authors followed the same measuring experimental protocol in the five twin units). Every single day, a different unit was tested, which included the presentation of 40 different gas conditions, presented in random order, exposing each unit to 10 concentration levels of Ethanol, Methane, Ethylene, and Carbon Monoxide (four classes). The conductivity of each sensor for 600 s in each experiment was acquired by using a sample rate of 100 Hz. The authors tested the detections platforms for 22 days, but only 16 days, the measurements were collected. Hence, the complete dataset comprises 640 records [15].

2.2 Deep Learning Models

The models generated were implemented using the *Python* programming language. In this study, three DL architecture implementations types were used for the classification tasks. The first type was a set of three simple DL models named SniffNets in [16]. The SniffNets were implemented employing the machine learning framework: Keras [17]. The second architecture implementation type was a DL model to perform meta-learning, adjusting the connections between different computing cells by differentiable search to obtain the best graph configuration while training. The authors called that methodology as Differentiable Architecture Search (DARTS) [18]. The DARTS implementation has been made available by its authors, but we adapted it so that the model could fit the shape of the target data. This implementation was created using the PyTorch library [19]. Finally, the third model corresponds to a simple Deep MLP model with only fully connected layers based on the model used in [7].

The input format used for the models with convolutional layers was a feature matrix with dimensions $R_f$ x $C_f$, where $R_f$ corresponds to the rows (represents the time interval) and the columns $C_f$ that corresponds to the gas sensors used to detect the specimens.

*2.2.1 SniffNets*

We generated three models with different architecture implementations based on [16], adapting the architectures to test the proposed rapid detection approach [7]. In the three models, we used the *softmax* as the activation function for the output layer. The code used in this study is available at [20]

The first model is a convolutional network named Sniff ConvNet in this document, which consists of two layers that apply a bi-dimensional convolution (Conv2D), followed by two fully connected (FC) layers. We used the activation function *ReLU* in both Conv2D and FC layers. The second model is a residual network named Sniff ResNet composed of two residual blocks, each one with two Conv2D layers. In each block, the first convolutional layer has a skip connection joined to the second convolutional layer output. Two FC layers follow the two residual blocks. Like as the Sniff ConvNet model, we used the activation function *ReLU* in the Conv2D and FC layers. The third model is a fusion neural network called Sniff Multinose. In this case, we adopted a different approach, where the feature matrix has a shape $R_f$ X $C_f$. We split the feature matrix by columns $C_f$, and each column was used as an input of a Multilayer Perceptron (MLP) model. Then, we concatenated the outputs of all MLP models and utilized them as inputs of another MLP network to complete the classification model.

*2.2.2 DARTS: Differentiable Architecture Search*

Searching for optimal neural network architectures is a task that can be both difficult and time consuming for the researchers. Seeking a way to automate this task, a study was proposed to perform this search using differentiation [18]. The key reason behind the use of differentiation is that the search space is continuous, and the algorithm performs the architecture optimization over the validation set performance. The algorithm performs the search in a network considered as a directed acyclic graph. Each node xi represents the output of a subnetwork in the chart. For example, xi can be a feature vector from a fully connected Multilayer perceptron or a feature map from a convolutional layer. Let O be the set of all the arcs(i,j) being an operation between the i-th node to the j-th node pondered by a factor α(i,j). The arc(i, j) represents the connection between nodes xi and xj. In which this connection is the o(i,j) operation which inputs are Xi and outputs Xj . After the initialization of a set of candidate operations between all nodes (i, j) of the graph. The search task is then performed by first computing the gradient of the loss function with respect to the factors α(i,j) and then concerning the weights of the model. Thus, after computing the minimal loss concerning the α and the weights in the arcs between (i,j), the algorithm determines the optimal architecture according to the values of α [18]. The code used in this study is available at [21].

*2.2.3 Deep MLP model*

We also used a Deep MLP model presented in [7]. The configuration of the model consists of eight layers each with *ReLU* as the activation function except for the output layer, in which we used *softmax*. In this model, the input layer was configured to have 100 neurons and all the hidden layers to have 30 neurons. The code used in this study is available at [20].

*2.2.4 Training Configurations*

We trained the three sets of DL models until to reach 20 epochs by using the Stochastic Gradient Descent (SGD) algorithm for optimization with a learning rate of 0.001 and a momentum of 0.9. Besides, we used the loss function called categorical cross-entropy and the holdout cross-validation method.

## 2.2.5 Configurations for the SVM model

The SVM model available in the scikit-learn library was used [22]. Furthermore, we defined the following parameters to optimize the model: A Radial Bayes Function (RBF) as the kernel, the regularization parameter C as 10, and the other settings as the default value. Given a dataset *D* with vectors of *n* features, the value computed for the *gamma* parameter is $(n \cdot variance(D_{flat}))^{-1}$. Where $variance(D_{flat})$ is the variance over the flattened dataset. The algorithm computed the *gamma* value over the normalized data, using standardization or *z-score normalization*.

## 3. Results and Discussion

According to the rapid detection approach, the rising window protocol was applied to find the early portion with the best validation accuracy (less error rate) in each dataset. We used different methods: the DARTS search model architecture, a deep MLP, three DL models called Sniff (ConvNet, Resnet, and Multinose), and SVM, to validate the proposed approach and determine if it could be applied independently to the classification method. The results of the experiments were summarized in **Table 1** to compare the performance on those windows (chosen windows and the last windows). The results let infer that the models generated with initials windows are similar or even outperform in the majority of cases the models obtained using the complete information of the measurements. The accuracy of the test data in the chosen windows is depicted in **Fig. 1**, to facilitate the visualization.

The chosen windows (with the best accuracy on test data) for each dataset are depicted in **Fig. 2**. It is important to remark that the first-window (w1) corresponds to the first 10% of measurement data, the second-window (w2) has 20% of the information, continuing until the ten-window (w10) that has the 100% of measurements data. The results suggest that, usually, using 30% or even less of information lets obtain suitable models. The necessary time to get the appropriate models for each dataset and classification method is detailed in **Table 2**. Those times are only a reference to know which methods work faster in the training process when is adopted the rapid detection approach.

The results showed similar accuracy for each dataset comparing the tested classification methods. The main difference is presented in the size of the best window. Comparing the Sniff models (gray bars in the figures), SniffMultinose reached the best-combined performance (15 datasets.) The DARTS algorithm (red bars in the figures) generates models that usually are similar to the models created by the Sniff architectures. Still, it is the method that needs more time in the training process to generate reliable models.

**Table 1.** Summary of experiments, showing the test data accuracy of the best window (b-win) and the accuracy of the last window (l-win), i.e., using all information after the gas injection for all tested datasets. For each dataset (rows), the best accuracy is highlighted.

| Dataset | Method | Sniff-ConvNet | | Sinff-Resnet | | Sniff-Multinose | | DARTS | | MLP | | SVM | |
|---|---|---|---|---|---|---|---|---|---|---|---|---|---|
| | | b-win | l-win | b-win | l-win | b-win | l-win | b-win | l-win | b-win | l-win | b-win | l-win |
| Wines | | 97.5 | 97.5 | 97.5 | 97.5 | 99.1 | 93.2 | 100 | 87 | 100 | 95.7 | 100 | 100 |
| | | w3 | w10 | w3 | w10 | w4 | w10 | w1 | w10 | w4 | w10 | w2 | w10 |
| Wines & ethanol | | 98 | 95.3 | 94 | 94 | 96.7 | 94.7 | 98.3 | 95 | 100 | 95 | 100 | 98.3 |
| | | w2 | w10 | w10 | w10 | w4 | w10 | w9 | w10 | w2 | w10 | w1 | w10 |
| Coffee | | 79.3 | 65.5 | 79.3 | 44.8 | 89.7 | 69 | 92 | 67 | 91.6 | 66.6 | 100 | 100 |
| | | w1 | w10 | w8 | w10 | w3 | w10 | w3 | w10 | w5 | w10 | w1 | w10 |
| Wind tunnelL1 | | 79.6 | 79.6 | 85.4 | 69.9 | 92.2 | 91.3 | 96.77 | 95.16 | 82.9 | 73.2 | 90.2 | 73.2 |
| | | w2 | w10 | w2 | w10 | w5 | w10 | w5 | w10 | w5 | w10 | w2 | w10 |
| Wind tunnelL2 | | 95.8 | 83.2 | 97.9 | 81 | 97.9 | 92.6 | 98.4 | 98.4 | 92.1 | 78.9 | 97.4 | 89.5 |
| | | w4 | w10 | w3 | w10 | w2 | w10 | w7 | w10 | w6 | w10 | w6 | w10 |
| Wind tunnelL3 | | 96.1 | 90.3 | 92.2 | 87.4 | 99 | 94.2 | 98.4 | 98.4 | 90.2 | 92.7 | 100 | 97.6 |
| | | w6 | w10 | w7 | w10 | w4 | w10 | w7 | w10 | w3 | w10 | w5 | w10 |
| Wind tunnelL4 | | 91.2 | 95.1 | 96.1 | 81.4 | 100 | 96.1 | 98.4 | 98.4 | 95.1 | 87.8 | 97.6 | 95.1 |
| | | w9 | w10 | w6 | w10 | w9 | w10 | w7 | w10 | w7 | w10 | w2 | w10 |
| Wind tunnelL5 | | 94.2 | 91.3 | 89.3 | 85.4 | 93.2 | 98 | 98.4 | 98.4 | 90.2 | 87.8 | 95.1 | 80.5 |
| | | w7 | w10 | w5 | w10 | w6 | w10 | w7 | w10 | w4 | w10 | w5 | w10 |
| Wind tunnelL6 | | 95.7 | 88.2 | 89.2 | 90.3 | 100 | 95.7 | 98.4 | 98.4 | 91.9 | 89.2 | 94.6 | 94.6 |
| | | w4 | w10 | w6 | w10 | w2 | w10 | w7 | w10 | w8 | w10 | w4 | w10 |
| Turbulent gas mixtures | | 60 | 52.2 | 53.3 | 34.4 | 57.7 | 30 | 73.3 | 60.1 | 72.2 | 72.2 | 77.7 | 86.1 |
| | | w3 | w10 | w4 | w10 | w7 | w10 | w3 | w10 | w10 | w10 | w4 | w10 |
| Twin gas sensorB1 | | 100 | 97.5 | 97.5 | 97.5 | 97.5 | 97.5 | 100 | 93 | 100 | 100 | 100 | 100 |
| | | w2 | w10 | w7 | w10 | w10 | w10 | w9 | w10 | w3 | w10 | w3 | w10 |
| Twin gas sensorB2 | | 97.5 | 91.2 | 96.2 | 100 | 100 | 80 | 96.9 | 92.1 | 100 | 90.6 | 100 | 96.9 |
| | | w3 | w10 | w6 | w10 | w1 | w10 | w10 | w10 | w1 | w10 | w1 | w10 |
| Twin gas sensorB3 | | 100 | 92.5 | 100 | 90 | 98.7 | 88.7 | 96.9 | 94.4 | 100 | 93.7 | 100 | 100 |
| | | w8 | w10 | w8 | w10 | w2 | w10 | w10 | w10 | w3 | w10 | w2 | w10 |
| Twin gas sensorB4 | | 92.5 | 52.5 | 100 | 92.5 | 100 | 90 | 87.5 | 77.6 | 100 | 62.5 | 100 | 100 |
| | | w3 | w10 | w8 | w10 | w3 | w10 | w9 | w10 | w5 | w10 | w2 | w10 |
| Twin gas sensorB5 | | 100 | 77.5 | 100 | 92.5 | 100 | 87.5 | 100 | 80.1 | 100 | 87.5 | 100 | 100 |
| | | w3 | w10 | w9 | w10 | w1 | w10 | w10 | w10 | w1 | w10 | w1 | w10 |

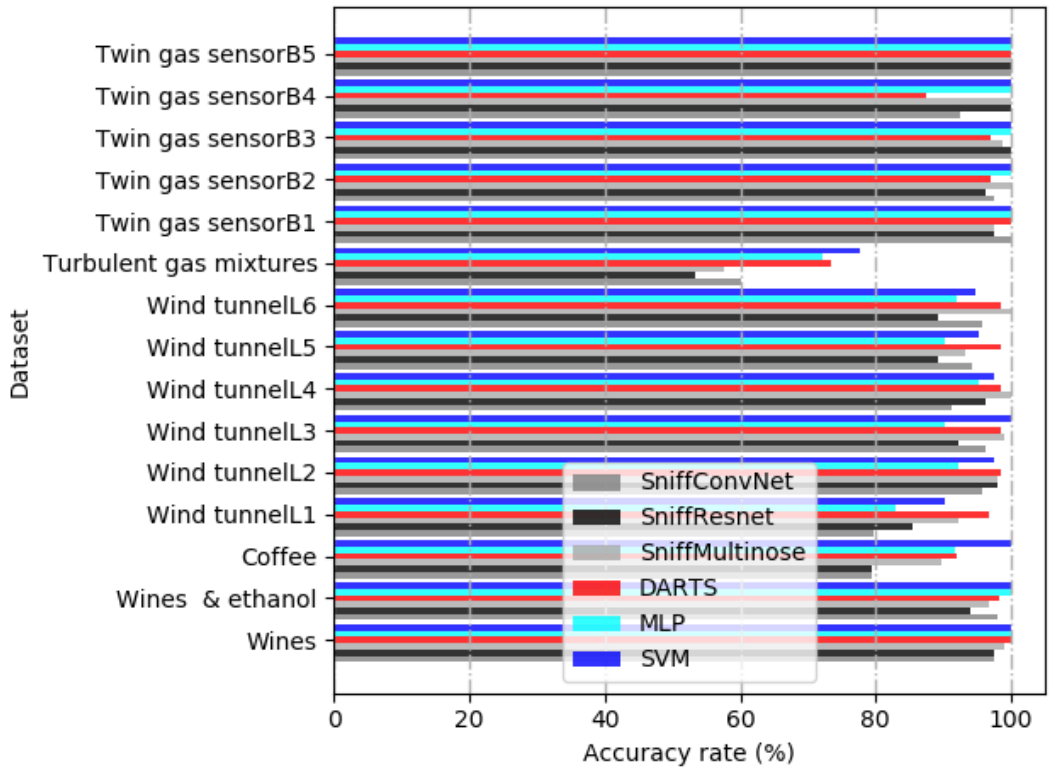

**Fig. 1.** The classification accuracy rate of the test data over the 15 datasets for the tested methods.

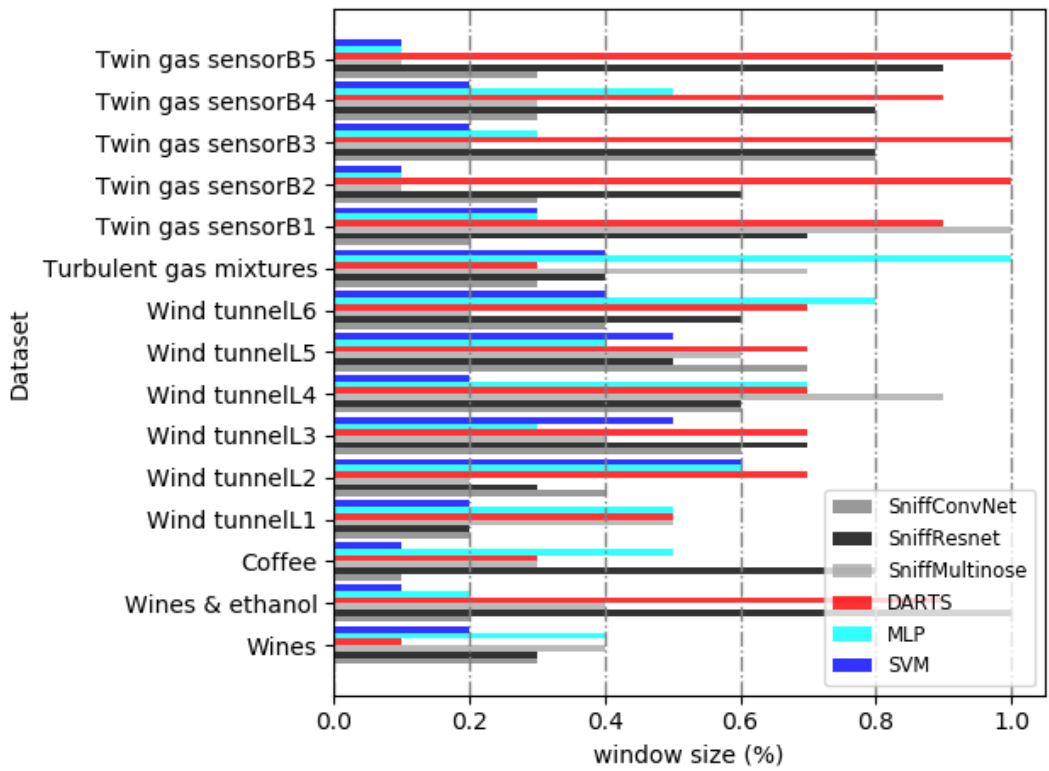

**Fig. 2.** The window size in percentage regarding the complete measurements over the 15 datasets for the tested methods.

**Table 2.** Training time in seconds for the tested classifications methods on the 15 datasets, only for reference.

| Dataset | Sniff-ConvNet | Sinff-Resnet | Sniff-Multinose | DARTS | MLP | SVM |
|---|---|---|---|---|---|---|
| Wines | 1475.2 | 148.93 | 83.55 | 5940 | 23.63 | 0.98 |
| Wines & ethanol | 1885.66 | 191.22 | 114.2 | 7380 | 33.28 | 1.65 |
| Coffee | 237.16 | 134.23 | 39.74 | 960 | 99.9 | 0.0375 |
| Wind tunnelL1 | 509.87 | 143.09 | 328.52 | 3360 | 65.676 | 6.22 |
| Wind tunnelL2 | 492.57 | 143.75 | 428.69 | 3120 | 71.746 | 3.44 |
| Wind tunnelL3 | 527.76 | 162.31 | 556.22 | 3360 | 82.245 | 5.65 |
| Wind tunnelL4 | 546.04 | 173.22 | 669.41 | 3300 | 90.31 | 5.82 |
| Wind tunnelL5 | 560.77 | 184.49 | 801.69 | 3300 | 99.32 | 6.17 |
| Wind tunnelL6 | 545.27 | 185.97 | 285.59 | 2820 | 104.273 | 5.11 |
| Turbulent gas mixtures | 3119.55 | 105.13 | 115.94 | 3120 | 55.734 | 2.83 |
| Twin gas sensorB1 | 5137.12 | 453.2 | 105.99 | 3300 | 28.692 | 1.1 |
| Twin gas sensorB2 | 5035.33 | 79.2 | 124.11 | 3360 | 24.93 | 1.2 |
| Twin gas sensorB3 | 5027.64 | 80.42 | 110.2 | 3360 | 31.236 | 1.1 |
| Twin gas sensorB4 | 2588.93 | 67.41 | 68.47 | 1800 | 32.4921 | 0.4 |
| Twin gas sensorB5 | 2574.06 | 79.88 | 84.79 | 1860 | 40.126 | 0.4 |

Analyzing the window size with the best accuracy on the test data, we conclude that based on the tested classification methods (DL techniques and SVM), the rapid detection approach is a reliable option to apply in electronic nose applications. Their results let to validate this methodology to be used in E-Nose datasets. Besides, it is relevant to remark that in the majority of cases, the SVM (blue bars in the figures) generates models that use an early portion of information with minor size **Fig. 2**, allowing to make forecasts in less time regarding the tested architectures. Additionally, the training time is quite less, reducing the computational cost. Therefore, the mentioned findings suggest that in the electronic nose field not worth it to use Convolutional Neural Networks (CNN) and deep learning techniques for classification tasks. These techniques increase the necessary time to generate reliable models and generally not reach much better results.

## 4. Conclusions

In this research, we validated the rapid detection approach [7] in several datasets with diverse electronic nose settings, showing that it is suitable to be used in this field, with better or similar accuracy compared against a conventional approach that needs the complete information of the measurements.

The investigation allowed finding that in the majority of cases is possible to obtain a reliable forecast using only the first 30% (even less) of the measure after the gas injection started. Therefore, subsequent investigations could focus on generating models using only this portion of the gas sensors signals, which entails reducing the time to produce models and make the forecasts (accelerating response).

In this work, we validated the proposed approach using several classification methods; the SVM algorithm and three different DL architecture approach: (i) the Differentiable Architecture Search (DARTS) algorithm, (ii) three deep learning models based on SniffNets, and (iii) a Deep MLP. Although deep learning models are useful when there is a large volume of data, and it can automatically identify patterns. The results showed that using SVM models in the majority of cases, the results are similar or even better and were consistent concerning the early portion of signals needed to make reliable forecasts. Therefore, SVM still is an excellent option in the electronic nose field and could be used to apply the rapid detection approach, as well, the tested deep learning techniques. Still, SVM needs less time for the training process against the other tested classifications methods.

**Acknowledgments**

We gratefully acknowledge the support of the following Brazilian agencies: CAPES, CNPq, and FACEPE, who promoted and financed our research project.